\documentclass[
    ,final            % use final for the camera ready runs
  ]
  {aipproc}

\layoutstyle{6x9}

\newcommand\etal{et al.}

\begin{document}

\title{Tokyo Axion Helioscope}


\author{Makoto Minowa}{
  address={Department of Physics and RESCEU, the University of Tokyo, 7-3-1 Hongo, Bunkyo-ku, Tokyo 113-0033, Japan}
}

%\author{<author2>}{
%  address={<common address for author2 and author3>}
%}

%\author{<author3>}{
%  address={<common address for author2 and author3>}
%  ,altaddress={<author1 address>} % additional visiting address
%}

\begin{abstract}
The idea of a magnetic axion helioscope was first proposed by Pierre Sikivie in 1983.
Tokyo axion helioscope was built exploiting its detection principle 
with a dedicated cryogen-free superconducting magnet
and PIN photodiodes for x-ray detectors. 
Solar axions, if exist, would be converted into x-ray photons
in the magnetic field. 
Conversion is coherently enhanced even for
massive axions by filling the conversion region with helium gas.
Its start up, search results so far and prospects are presented. 
\end{abstract}

\maketitle

%%%%%%%%%%%%%%%%%%%%%%%%%%%%%%%%%%%%%%%%%%%%
%% MAINMATTER
%%%%%%%%%%%%%%%%%%%%%%%%%%%%%%%%%%%%%%%%%%%%

\section{Solar axion detection, the principle}

Existence of the axion is implied
to solve the strong CP problem of Quantum chromodynamics (QCD)\cite{RDP,SW,FW,JEK}.  
The axion would be produced
in the solar core through the Primakoff effect. 
It can be converted back to an x-ray in a strong
magnetic field in the laboratory by the inverse process. 
This is the principle of the solar axion detection by the axion helioscope and
appeared for the first time in  the paper\cite{PS} entitled 
"Experimental Tests of the 'invisible' axion" 
written by Pierre Sikivie in 1983, 27 years ago.
Sikivie indeed mentioned the idea of the axion helioscope in this paper.

The conversion process is coherent when the axion and photon remain in phase over the length of
the magnetic field\cite{R-S} if the axion is almost massless.
Coherence can be maintained even for higher mass axions if the conversion region is filled with
a low-Z buffer gas like helium\cite{KVB,R-S}.

\section{Experimental efforts}
A possible laboratory detector design was first proposed by K. van Bibber \etal\cite{KVB}, and 
there was a pioneering helioscope experiment by Lazarus \etal\cite{Lazarus} in 1992. 
It was the first solar axion search experiment using magnetic field in a laboratory.
At that time, however, a fixed magnet was used and running time was short.

They put significant upper limits on $g_{a\gamma\gamma}$ for the axion mass less than 0.11\, eV.
The measurements were done with vacuum magnetic field region for the lower axion mass region and with
a buffer helium gas of up to 100\,Torr for the higher mass region.

We started the construction of Tokyo axion helioscope in 1995, and the phase 1 result\cite{sumico1997} was
published in 1998 without a buffer gas in the manetic field region.
Then, phase 2 measurement for the axion mass less than 0.27\, eV 
with a helium gas was made and result was published
in 2002\cite{sumico2000}.

In the mean time, CAST experiment\cite{CAST} was getting started in CERN and published their result for measurements with vacuum
magnetic field region.

The third phase of Tokyo axion helioscope was then prepared, 
and the result was published for the axion mass region 
between 0.84\, eV and 1.00\, eV in 2008\cite{sumico2008}.
It is the first result to search for the axion in the $g_{a\gamma\gamma}$-$m_a$ parameter region of the
preferred QCD axion models with a magnetic helioscope.

Almost at the same time, CAST published their new results\cite{CAST2} with a helium-4 buffer gas 
in the magnetic field region.

\section{Tokyo Axion Helioscope}
\begin{figure}[hb]
  \includegraphics[height=.3\textheight]{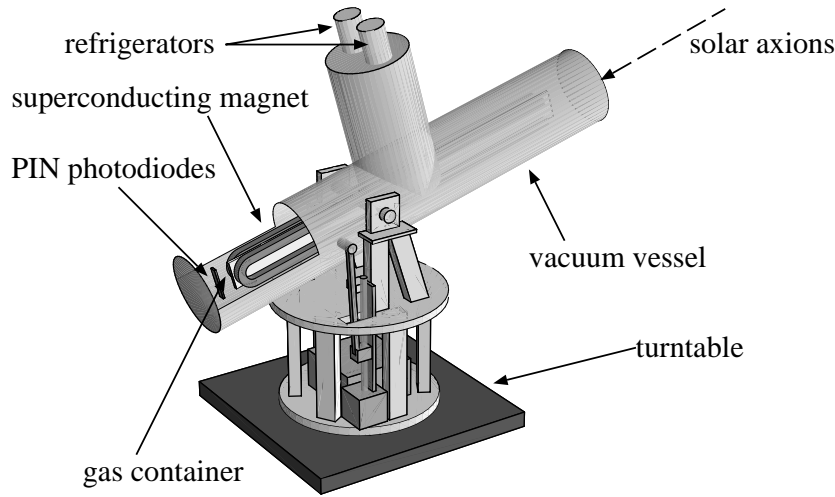}
  \caption{Tokyo axion helioscope}
  \label{fig:Sumico}
\end{figure}

The schematic view of the Tokyo axion helioscope is shown in Fig. \ref{fig:Sumico}. 
Its main components are
common to phase 1 through 3 of the Tokyo
axion helioscope experiments\cite{sumico1997, sumico2000, sumico2008}. 
It is designed to track the sun in
order to achieve long exposure time. It consists of a dedicated superconducting magnet, X-ray detectors,
a gas container, and an altazimuth mounting.

%%%%%%%%%%%%%%%%%%%%%%%%%%%%%%%%%%%%%%%%%%%%
%% Sample figure:
%%
%% The option [height=...] scales the picture to the given height,
%% without it it would be printed at its nominal size
%%%%%%%%%%%%%%%%%%%%%%%%%%%%%%%%%%%%%%%%%%%%

The superconducting magnet\cite{sato1997}
consists of two 2.3-m long race-track
shaped coils running parallel with a 20-
mm wide gap between them. 
The magnetic
field in the gap is 4\,T in the direction perpendicular
to the helioscope axis. 
The coils are kept
at 5 -- 6K during operation. 
The magnet
was made cryogen-free by making two
Gifford-McMahon refrigerators to cool it
directly by conduction, and is equipped
with a persistent current switch. 
Thanks
to these features, the magnet can be
freed from thick current leads after excitation,
and the magnetic field is very
stable for a long period of time without
supplying current.
The container to hold dispersion-matching buffer gas is inserted in the $20\times92\,\mathrm{mm^2}$ aperture of the
magnet. Its body is made of four 2.3-m long 0.8-mm thick stainless-steel square pipes welded
side by side to each other.
Sixteen PIN photodiodes, Hamamatsu Photonics S3590-06-SPL, are used as the X-ray detectors
\cite{naniwaPIN}, whose chip sizes are $11\times11\times0.5\rm\,mm^3$ each. 
The effective area of a photodiode was measured
formerly using a pencil-beam X-ray source, and found to be larger than $9\times9\,\mathrm{mm^2}$. 
It has an
inactive surface layer of $0.35\,\mu\mathrm{m}$ \cite{akimotoPIN}.
The entire axion detector is constructed in a vacuum vessel and the vessel is mounted on an
altazimuth mount. Its trackable altitude ranges from $-28^\circ$ to $+28^\circ$ and its azimuthal direction
is designed to be limited only by a limiter which prevents the helioscope from endless rotation.
However, in the present measurement, the azimuthal range is restricted to about 60$^\circ$ because
a cable handling system for its unmanned operation is not completed yet.

\section{Results}
\begin{figure}[hb]
  \includegraphics[height=.5\textheight]{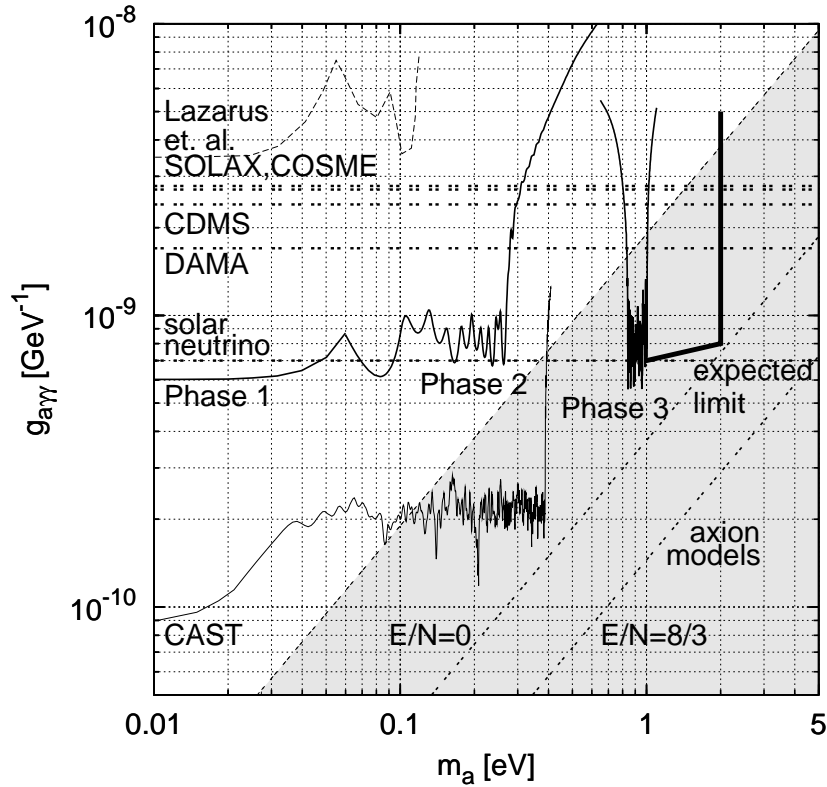}
  \caption{Exclusion plot on $g_{a\gamma\gamma}$ vs $m_a$.}
  \label{fig:exclusion}
\end{figure}

Fig.~\ref{fig:exclusion} shows
the upper limit to $g_{a\gamma\gamma}$ plotted as a function of $m_a$.
Our limits from the phase 1 through 3 of the Tokyo
axion helioscope experiments\cite{sumico1997, sumico2000, sumico2008}
and some other bounds are also
plotted in the same figure.

The SOLAX~\cite{solax1999}, COSME~\cite{cosme2002} and DAMA~\cite{DAMA2001} are solar axion experiments
which exploit the coherent conversion
on the crystalline planes \cite{Pascos} in a germanium and a NaI detector.
Also shown is the solar limit inferred from the
solar neutrino flux consideration\cite{Solar}.
The experiment by Lazarus \etal~\cite{Lazarus} and CAST~\cite{CAST, CAST2} 
are the same kind of helioscope experiments as mentioned above.
The latter utilizes large decommissioned magnets of the LHC at CERN.
Its limit is better than our previous limits by a factor of four to seven
due to its higher magnetic field and longer field region.
However, the upper limits around 1\, eV by Tokyo axion helioscope phase 3 measurement is a unique
result in this mass region.

\section{Prospects}
We are now preparing the search for solar axion with mass over 1\, eV introducing 
higher density helium gas than that of phase 3. 
Figure~\ref{fig:exclusion} shows the expected upper limit of next measurement. 

\begin{figure}[hb]
  \includegraphics[height=.3\textheight]{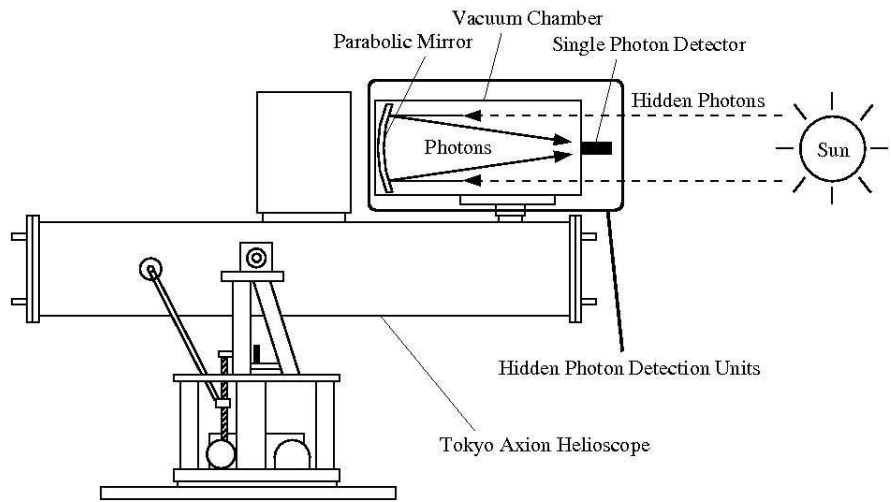}
  \caption{The schematic view of the apparatus
to search for hidden photons from the sun.}
  \label{fig:HP}
\end{figure}

We are also constructing an additional detection unit on the Tokyo axion helioscope
to search for hidden sector photons.
The hidden sector photon is another kind of weakly interacting particles. 
Existence of the hidden sector photon is predicted by several extensions of the standard model. 
If light hidden sector photons exist, they could be produced through kinetic
mixing with solar photons~\cite{Redondo:2008aa, Gninenko:2008pz}.
Therefore it is natural to consider the Sun as a source of low energy hidden photons. 
A detection schematics of hidden photons from the sun is shown in Fig.~\ref{fig:HP}. 
Unlike the case of the axion, no magnetic field is required to transform hidden sector photons into photons.

%%%%%%%%%%%%%%%%%%%%%%%%%%%%%%%%%%%%%%%%%%%%%%%%
%% BACKMATTER
%%%%%%%%%%%%%%%%%%%%%%%%%%%%%%%%%%%%%%%%%%%%%%%%

\begin{theacknowledgments}
The author thanks the former director general of KEK, Professor H. Sugawara, for his support
in the beginning of the helioscope experiment and all the collaboration members of Tokyo axion helioscope. 
This research was partially supported by the
Japanese Ministry of Education, Science, Sports and Culture, Grant-in-Aid for COE Research
and Grant-in-Aid for Scientific Research (B), and also by the Matsuo Foundation.
\end{theacknowledgments}

%%%%%%%%%%%%%%%%%%%%%%%%%%%%%%%%%%%%%%%%%%%%%%%%
%% The bibliography can be prepared using the BibTeX program or
%% manually.
%%
%% The code below assumes that BibTeX is used.  If the bibliography is
%% produced without BibTeX comment out the following lines and see the
%% aipguide.pdf for further information.
%%
%% For your convenience a manually coded example is appended
%% after the \end{document}
%%%%%%%%%%%%%%%%%%%%%%%%%%%%%%%%%%%%%%%%%%%%%%%%

\end{document}